\documentclass[reprint,
 superscriptaddress,
 amsmath,
 amssymb,
 aps,
 pra,
]{revtex4-1}
\usepackage[utf8]{inputenc}
\usepackage{graphicx}
\usepackage{dcolumn}
\usepackage{bm}
\usepackage{natbib}
\usepackage{dirtytalk}
\usepackage{csquotes}
\usepackage{subcaption}
\usepackage{multirow}
\usepackage{textgreek} 

\begin{document}

\title{Strong electron-electron interactions in a dilute weakly-localized metal near a metal-to-insulator transition}
\author{Nicolò D'Anna}
\email{ndanna@ucsd.edu}
\altaffiliation{Present address: Department of Physics
University of California San Diego, La Jolla, CA 92093, USA}
\affiliation{Paul Scherrer Institute, 5232 Villigen PSI, Switzerland}
\affiliation{Laboratory for Solid State Physics and Quantum Center, ETH Zurich, Zurich, Switzerland}
\author{Jamie Bragg}
\affiliation{London Centre for Nanotechnology, University College London, WC1H 0AH, London, UK}
\affiliation{Department of Electronic and Electrical Engineering, University College London, London WC1E 7SE, UK}
\author{Aidan~G.~McConnell}
\affiliation{Paul Scherrer Institute, 5232 Villigen PSI, Switzerland}
\affiliation{Laboratory for Solid State Physics and Quantum Center, ETH Zurich, Zurich, Switzerland}
\author{Jakub~Vonka}
\affiliation{Paul Scherrer Institute, 5232 Villigen PSI, Switzerland}
\author{Procopios~C.~Constantinou}
\affiliation{Paul Scherrer Institute, 5232 Villigen PSI, Switzerland}
\affiliation{London Centre for Nanotechnology, University College London, WC1H 0AH, London, UK}
\affiliation{Department of Physics and Astronomy, University College London, WC1E 6BT, London, UK}
\author{Juerong~Li}
\affiliation{Advanced Technology Institute, University of Surrey, Guildford, GU2 7XH UK}
\author{Taylor~J.Z.~Stock}
\affiliation{London Centre for Nanotechnology, University College London, WC1H 0AH, London, UK}
\affiliation{Department of Electronic and Electrical Engineering, University College London, London WC1E 7SE, UK}
\author{Steven~R.~Schofield}
\affiliation{London Centre for Nanotechnology, University College London, WC1H 0AH, London, UK}
\affiliation{Department of Physics and Astronomy, University College London, WC1E 6BT, London, UK}
\author{Neil~J.~Curson}
\affiliation{London Centre for Nanotechnology, University College London, WC1H 0AH, London, UK}
\affiliation{Department of Electronic and Electrical Engineering, University College London, London WC1E 7SE, UK}
\author{Y. Soh}
\affiliation{Paul Scherrer Institute, 5232 Villigen PSI, Switzerland}
\author{Marek~Bartkowiak}
\affiliation{Paul Scherrer Institute, 5232 Villigen PSI, Switzerland}
\author{Simon Gerber}
\affiliation{Paul Scherrer Institute, 5232 Villigen PSI, Switzerland}
\author{Markus Müller}
\affiliation{PSI Center for Scientific Computing and Data, Laboratory for Theoretical and Computational Physics, Paul Scherrer Institute, 5232 Villigen PSI, Switzerland}
\author{Guy Matmon}
\affiliation{Paul Scherrer Institute, 5232 Villigen PSI, Switzerland}
\author{Gabriel Aeppli}
\email{aepplig@ethz.ch}
\affiliation{Paul Scherrer Institute, 5232 Villigen PSI, Switzerland}
\affiliation{Laboratory for Solid State Physics and Quantum Center, ETH Zurich, Zurich, Switzerland}
\affiliation{Institute of Physics, EPF Lausanne, 1015 Lausanne, Switzerland}

\begin{abstract}
Because it is easily switched from insulator to metal either via chemical doping or electrical gating, silicon is at the core of modern information technology and remains a candidate platform for quantum computing. The metal-to-insulator transition in this material has therefore been one of the most studied phenomena in condensed matter physics, and has been revisited with considerable profit each time a new fabrication technology has been introduced. Here we take advantage of recent advances in creating ultra-thin layers of Bohr-atom-like dopants to realize the two-dimensional disordered Hubbard model at half-filling and its metal-to-insulator transition (MIT) as a function of mean distance between atoms.
We use gas-phase dosing of dopant precursor molecules on silicon to create arsenic and phosphorus $\delta$-layers as thin as 0.4~nm and as dilute as 10$^{13}$~cm$^{-2}$. 
On approaching the insulating state, the conventional weak localization effects, prevalent at high dopant densities and due to orbital motion of the electrons in the plane, become dominated by electron-electron interaction contributions which obey a paramagnetic Zeeman scaling law. The latter make a negative contribution to the conductance, and thus cannot be interpreted in terms of an emergent Kondo regime near the MIT. 
\end{abstract}
\maketitle

The physics of two-dimensional (2D) electron gases has been the subject of extensive experimental and theoretical research efforts since the 1960s, and has led to some of the most remarkable discoveries in solid-state physics, including the quantum Hall effect \cite{cage2012quantum} and topological quantum states of matter \cite{QSHI_HgTe}. Despite considerable advances, 2D materials remain the subject of intense ongoing research because of the many possible consequences of combined disorder, inter-site hopping of electrons, and Coulomb interaction effects. A problem that has proven to be elusively difficult to solve is the nature of the metal-to-insulator transition (MIT) in two-dimensions. It was widely thought to be non-existent in the 80s and 90s, because of predictions that localization would turn any 2D material into an insulator \cite{PhysRevLett.42.673}, but was later observed in most 2D systems \cite{Graphene_MIT_2011,VdW_MIT_2019,SiGe_2020,GaAs_1999,AlAs_2013,ZnO_2006,Zhang:2016aa}.
While the 2D MIT has been studied in many materials, the mechanisms driving it and the phases associated with it remain open questions \cite{SHASHKIN2021168542, kirk20192d, arovas2022hubbard}.
In silicon, 2D electron layers can be formed in a number of ways, for example with MOSFETs \cite{PhysRevB.50.8039}, quantum wells \cite{doi:10.1063/1.1639507}, and $\delta$-doped layers \cite{schubert1996delta}. In these systems the electron mobility and the disorder strength are substantially different, as visible in Table~\ref{Table_2DEGs}; nonetheless they all exhibit the MIT at similar interaction strength parameter $r_s \approx 20$ \cite{PhysRevB.99.081106}, defined as the ratio of the Coulomb energy and the Fermi energy ($r_s = g_\text{v}/(\pi n_\text{2D})^{1/2}a_\text{B}$), where $n_\text{2D}$ is the sheet electron density, $a_\text{B}$ the effective Bohr radius, and $g_\text{v}$ the valley degeneracy of the host material).
The critical value of $r_s$ depends on the disorder strength and is lower at high disorder \cite{Walsh_2019}.
There is no agreement on the mechanisms driving the MIT \cite{dolgopolov2019}, but a number of distinct insulating phases have been proposed for clean \cite{SHASHKIN2021168542} and disordered systems \cite{Yildiz_2009,DasSarma_crossover,DasSarma_perc}, differing mostly in their magnetic behavior \cite{potter2012quantum}.
\begin{table}[ht]
\small
\centering
\begin{tabular}{cccccc}
 \hline
2D system & $\ell$ [nm]& $\ell_\phi$ [nm]&  $n_\text{2D} [$10$^{14}$cm$^{-2}$] & $\delta$ [nm] & $r_s$\\
 \hline
 \textbf{Current work} &  \multirow{2}{*}{3}&  \multirow{2}{*}{70} & \multirow{2}{*}{0.12-1.9} & \multirow{2}{*}{0.4-1.8} & \multirow{2}{*}{1.4-5.5} \\
 \mbox{\textbf{Si:As $\&$ Si:P}} & & & & & \\
\mbox{\textbf{Ge:P}~\cite{PhysRevLett.112.236602}} & 20 & 400 & 0.3-1.4 & 1.49 & 0.6-1.3\\
Si MOSFET~\cite{MC_silicon_inversion_layers}& 48 & 252 & 0.4-0.8$\times$10$^{-4}$  & 10 & 10-20\\
QW 2DHG~\cite{Shashkin_2005}& 26 & 61 & 0.4-3.8$\times$10$^{-4}$ & 0.5 & 10-20\\
Graphene~\cite{PhysRevLett.107.166602}& 50 & 900 & 0.005 & 0.5-1 & $\sim$ 6-30 \\
 \hline
\end{tabular}
\caption{\textbf{Characteristics of various 2D electron gases.} The systems in bold are half-filled. $\ell$ is the mean free path, $\ell_\phi$ the coherence length, $n_\text{2D}$ the 2D sheet electron density, $\delta$ the electron layer thickness and $r_s$ the interaction strength.}
\label{Table_2DEGs}
\end{table}

 \begin{figure*}[ht] 
 \centering   
  \begin{subfigure}[b]{0.43\textwidth}  
  \includegraphics[width=\textwidth]{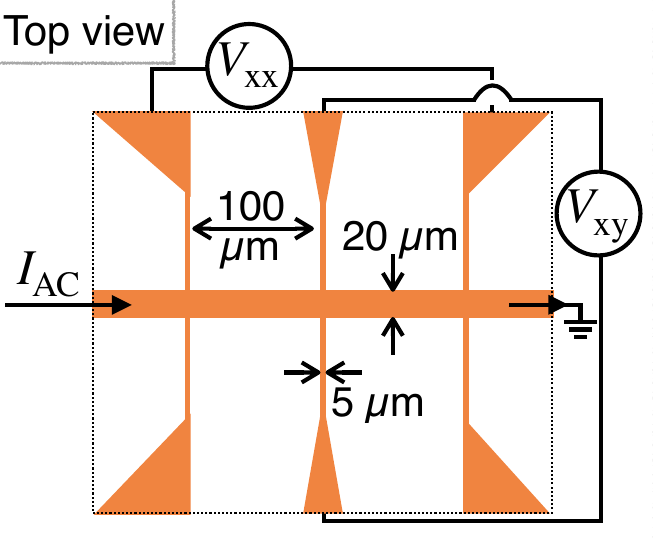}
  \caption{\label{fig_setup_a}}
  \end{subfigure}
\hfill
  \begin{subfigure}[b]{0.43\textwidth}  
  \includegraphics[width=\textwidth]{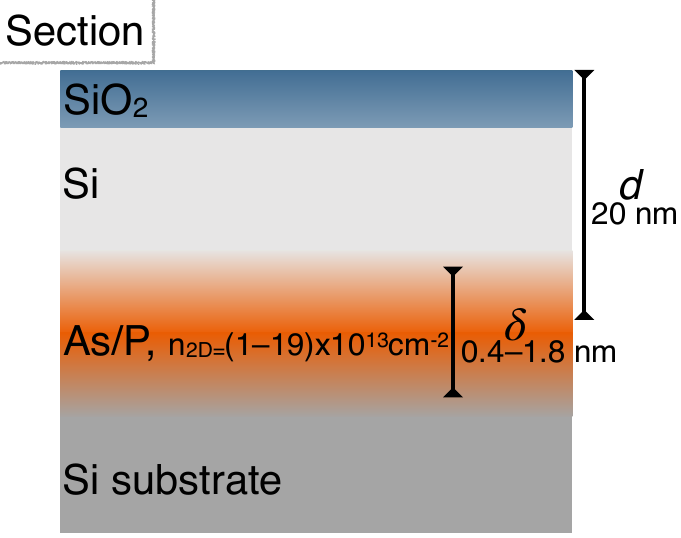}
  \caption{\label{fig_setup_b}}
  \end{subfigure}
 \caption{\textbf{Experimental setup.} 
 	(a) Hall-bar geometry of the contacts allowing for 4-point resistance measurements. (b) Schematic of the sample composition. The donor layer is encapsulated in the silicon lattice at a depth $d \sim$~20~nm below the surface, covering a width $\delta$ between 0.4~nm and 1.8~nm having a 2D electron density $n_\text{2D} = (1$–$19)\times10^{13}$cm$^{-2}$.
    }
\label{figsample}            
\end{figure*}

Here we report on the low temperature magneto-conductance of arsenic- and phosphorus-doped silicon $\delta$-layers at various doping densities near the MIT. Owing to the unprecedented combination of high densities and thinness for such $\delta$-layers in silicon (see Table.~\ref{Table_2DEGs}), we are able to access and highlight a new regime dominated by strong Coulomb interactions, where the interaction parameter is in the range $r_s = 1.4-5.5$ for an unusually high density of carriers.
In this system the number of free electrons corresponds to the number of donors, as confirmed by the consistency between  Hall effect and X-ray fluorescence data~\cite{SiAs_Xray_fluorescence}. It should thus be well described by a highly disordered half-filled 2D Hubbard model. 
Recent theoretical studies of the ordered half-filled 2D Hubbard model \cite{Walsh_2019} predict that at low temperature and low interaction strength there is a first order transition from the metallic to the insulating phase, as was already suggested by Mott  in the site-ordered Hubbard model with no disorder~\cite{mott2004metal}. As usual, the discontinuous  transition comes with a coexistence region.

Following general arguments~\cite{bergman1977critical}, the addition of disorder will render the transition of percolative nature, being continuous in full equilibrium, while hysteretic behavior and a coexistence of various metastable states, that  differ in their local conduction properties, is expected to be inherited from the disorder-free limit.\\

\begin{figure*}[ht] 
  \begin{subfigure}{0.32\textwidth}  
  \includegraphics[width=\textwidth]{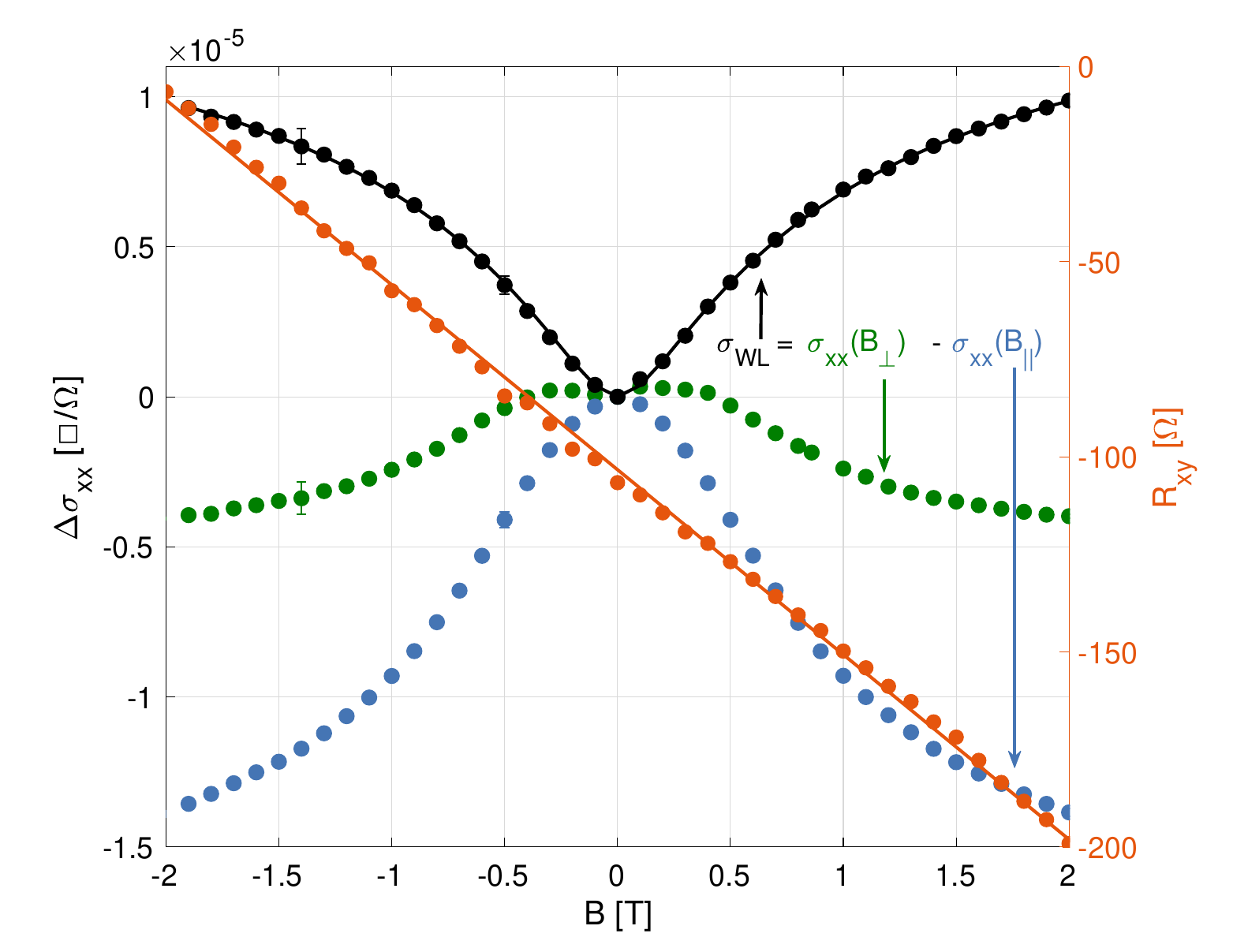}
  \caption{\label{fig_MR_a}}
  \end{subfigure}
\hfill     
  \begin{subfigure}{0.32\textwidth}  
  \includegraphics[width=\textwidth]{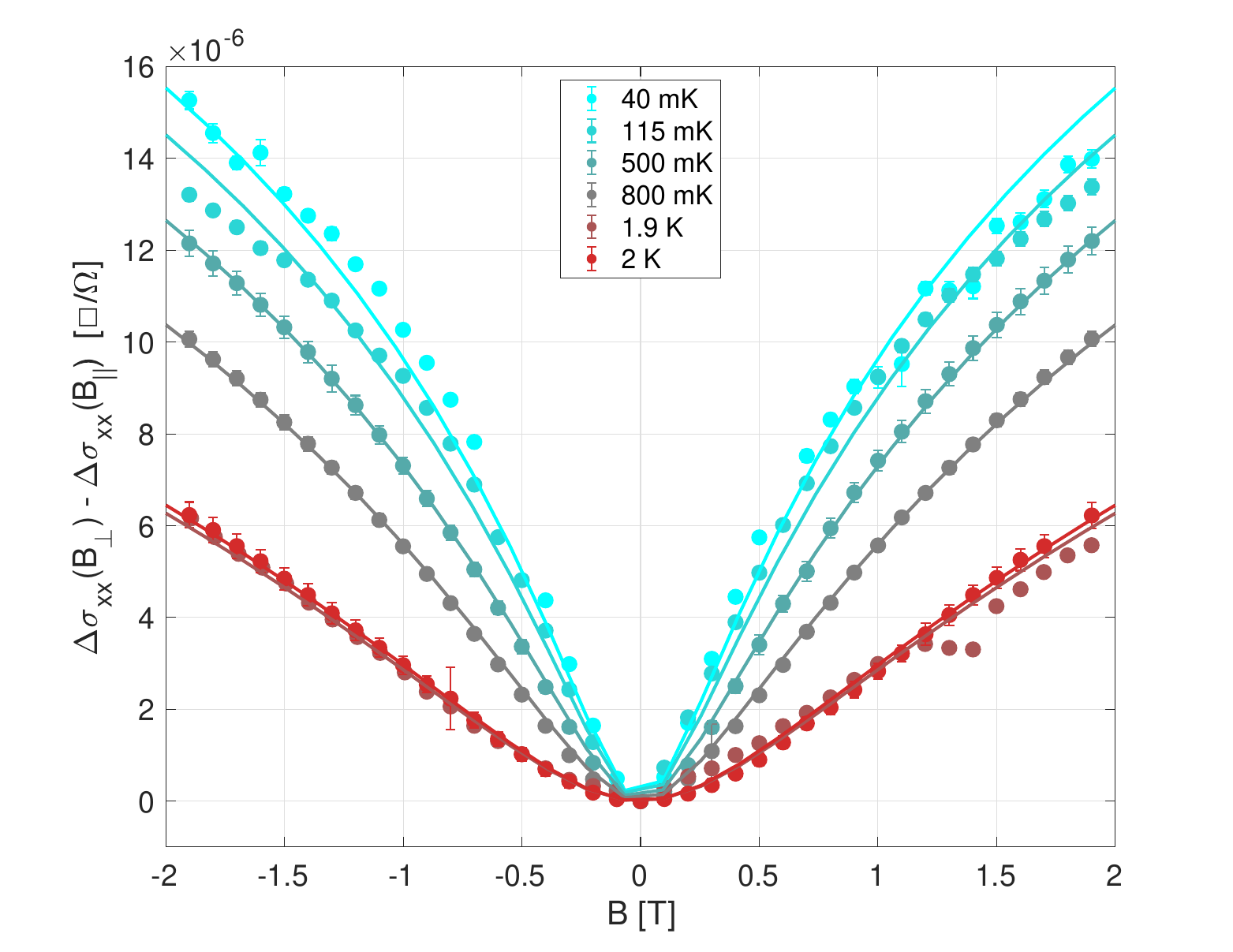}
  \caption{\label{WL_MR_vs_T}}
  \end{subfigure}
\hfill 
\begin{subfigure}{0.32\textwidth} 
  \includegraphics[width=\textwidth]{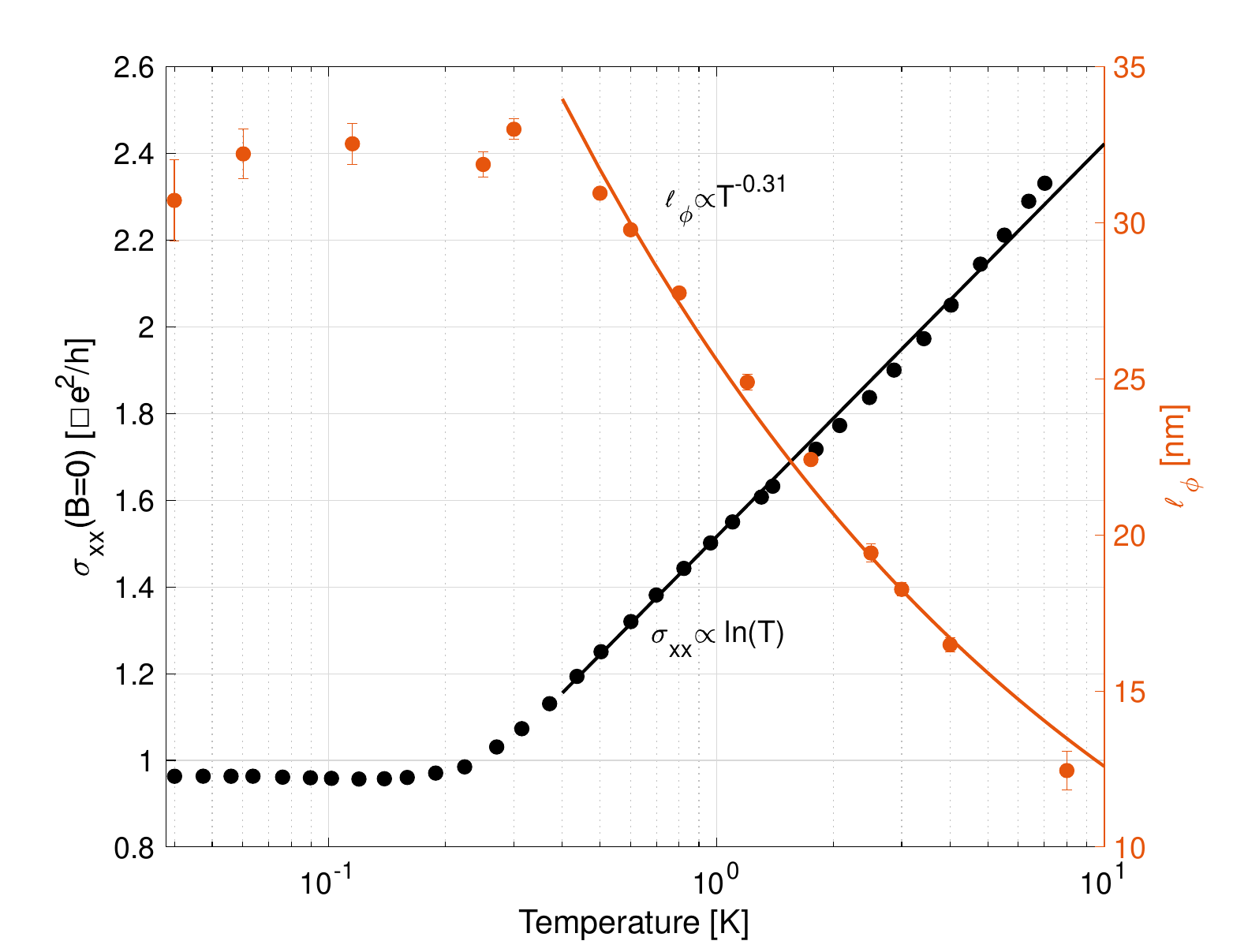}
  \caption{\label{Greenford_vs_T}}
  \end{subfigure}
 \caption{\textbf{Typical magneto-conductance data.} 
 	(a)  At temperature $T= 115 {\rm}$~mK, the magnetic field is set parallel to the plane of the $\delta$-layer (blue points - with no dependence on the relative orientation to the current) and perpendicular to it (green points). The black points are the difference between in- and out-of-plane magneto-conductance {per square}, the solid black line being a fit with weak-localization corrections, Eq.~(\ref{eq_HLN}) minus Eq.~(\ref{eq_Sullivan}). The orange data (right y-axis) is a Hall measurement ($R_{xy}$) taken at 1.75~K, fitted linearly by the solid line. 
	(b) Difference between out-of-plane and in-plane magneto-conductance {per square} for various temperatures. The solid lines are fits with Eq.~(\ref{eq_HLN}) minus Eq.~(\ref{eq_Sullivan}).
	(c) Conductance per square versus temperature in black, displaying a logarithmic dependence as expected for weak localization down to $T_\text{c}$ $\approx$ 300 mK. The coherence length $\ell_\phi$ obtained from the fits in (b) is shown in orange. It follows a power law $T^{-0.31}$ down to $T_\text{c}$.
	All data in (a), (b), and (c) are obtained from a sample of electron density $n = (1.18 \pm 0.01) \times 10^{13}$~cm$^{-2}$.}
\label{fig_MR}            
\end{figure*}

\noindent\textbf{Two-dimensional dopant layers in silicon}\\
The samples studied in this work are fabricated by gas-phase dosing of dopant precursor arsine (AsH$_3$) or phosphine (PH$_3$) molecules on a flat silicon (001) surface, followed by homoepitaxial overgrowth of crystalline silicon. These samples are then etched into Hall bars and contacted with aluminium \cite{Stock_As_count,d2022low}, as shown in Fig.~\ref{fig_setup_a}. Further details are in the supplemental information.
The samples are measured in a dilution refrigerator with a 2\,T 4$\pi$ vector magnet and 30~mK base temperature. To measure the longitudinal resistance $R_{xx}$ and the transverse resistance $R_{xy}$ a four-point scheme is used, where a source applies a constant low frequency current (13.7~Hz) and a lock-in amplifier measures the voltage drop between two inner contacts, such that the measured voltage is insensitive to the contact resistance. The current used for the measurements ranges between 10~pA and 1~nA, and is chosen to be well within the devices' linear IV response.
The $\delta$-layers have dopant densities ranging from 1$\times$10$^{13}$ to 2$\times$10$^{14}$~cm$^{-2}$, located at depths between 15 and 30~nm below the surface (Fig.~\ref{fig_setup_b}). The arsenic layers have thicknesses between 0.4 and 1.8~nm, as determined by magneto-conductance measurements.
The measured values for all samples are listed in the supplemental section~\ref{sect_supplemental} in Table.~\ref{table_samples}.
At these dopant densities the $\delta$-layers are still metallic, but very close to the metal-insulator transition, which is estimated to occur at $1\times10^{13}$~cm$^{-2}$ \cite{PhysRevB.87.125411, siP_sheet}. The parameter $r_s$, which quantifies the  strength of interactions, increases from 1.4 in the dense layers to 5.5 in the dilute layers, such that electron-electron interaction effects become non-negligible.

The temperature dependence of the samples' longitudinal conductivity ($\sigma_{xx} \equiv 1/R_{xx}\times L/W $ with $L$ and $W$ the sample length and width, respectively) in the absence of magnetic field is shown in Fig.~\ref{Greenford_vs_T}. It exhibits two regimes. The conductivity decreases logarithmically with decreasing temperature $T$ down to about $T_\text{c} = 300$ mK, with a dependence $\sigma_{xx}\propto\ln{T}$, and saturates at lower temperatures. Such a temperature dependence has been observed in many different materials \cite{2_to_3_crossover, Polley:2011aa, CHEUNG19941225, FOURNIER20001941, IHN20081851, PhysRevB.78.125409}, but the origin and intrinsic nature of the saturation remains debated \cite{RevModPhys.91.011002}.
At temperatures above the saturation of the conductivity, the electronic system is a conventional weakly-localized two-dimensional conductor with strong electron-electron interactions.\\

\noindent\textbf{Conventional weakly-localized regime}\\
In the two-dimensional dopant layers studied here, the conduction at temperatures above the crossover temperature $T_\text{c}$ is characterized by a logarithmic temperature dependence of the conductivity which remains larger than the quantum of conductance $G_0 = e^2/h$, indicative of diffusive transport, for which weak-localization (WL) effects can be expected \cite{PhysRevB.22.5142}. 
Weak localization is a quantum correction to the Drude conductivity that stems from the interference of diffusive electrons with themselves through self-intercepting loops. For weak localization to arise the electron coherence length $\ell_\phi$ must exceed the mean free path $\ell$, which sets the scale of the smallest loops. Weak localization is strongest if time reversal symmetry is not broken; when a magnetic field is applied, the time-reversal symmetry is broken as the field couples to the electron's orbital motion. For a two-dimensional conduction layer with the field perpendicular to the layer $B_{\perp}$, the ensuing change in conductivity is given by the following expression derived by Hikami, Larkin and Nagaoka \cite{HLN}:
\begin{equation}
\Delta\sigma(B_{\perp}) = \frac{G_{0}}{\pi} \left[  \psi\left(\frac{1}{2} +\frac{B_\phi}{B_{\perp}} \right) -  \psi\left(\frac{1}{2} +\frac{B_\ell}{B_{\perp}} \right) +\ln\left(\frac{2\ell_{\phi}^2}{\ell^2}\right) \right],
\label{eq_HLN}
\end{equation}
where $\psi(x)$ is the digamma function. The phase breaking field is given by $B_\phi = \frac{\hbar}{4e \ell_{\phi}^2}$ and the elastic characteristic field by $B_\ell=\frac{\hbar}{2e\ell^2}$. 
 If the layer has a finite thickness $\delta < \ell_\phi$, a magnetic field applied in the conduction plane $B_{\parallel}$ still has a small orbital effect 
\begin{equation}
\Delta\sigma(B_{\parallel}) = \frac{G_0}{\pi}\ln(1+\gamma B_{\parallel}^2),
\label{eq_Sullivan}
\end{equation}
that depends on the thickness of the layer through the single parameter $\gamma = \delta^2\sqrt{\frac{4\pi}{n}}\left(\frac{e}{\hbar}\frac{\ell_\phi}{\sqrt{\ell}}\right)^2$ \cite{sullivan,in_plane_MC}.
\\

\begin{figure*}[ht] 
  \centering
  \begin{subfigure}{0.43\textwidth}  
  \includegraphics[width=\textwidth]{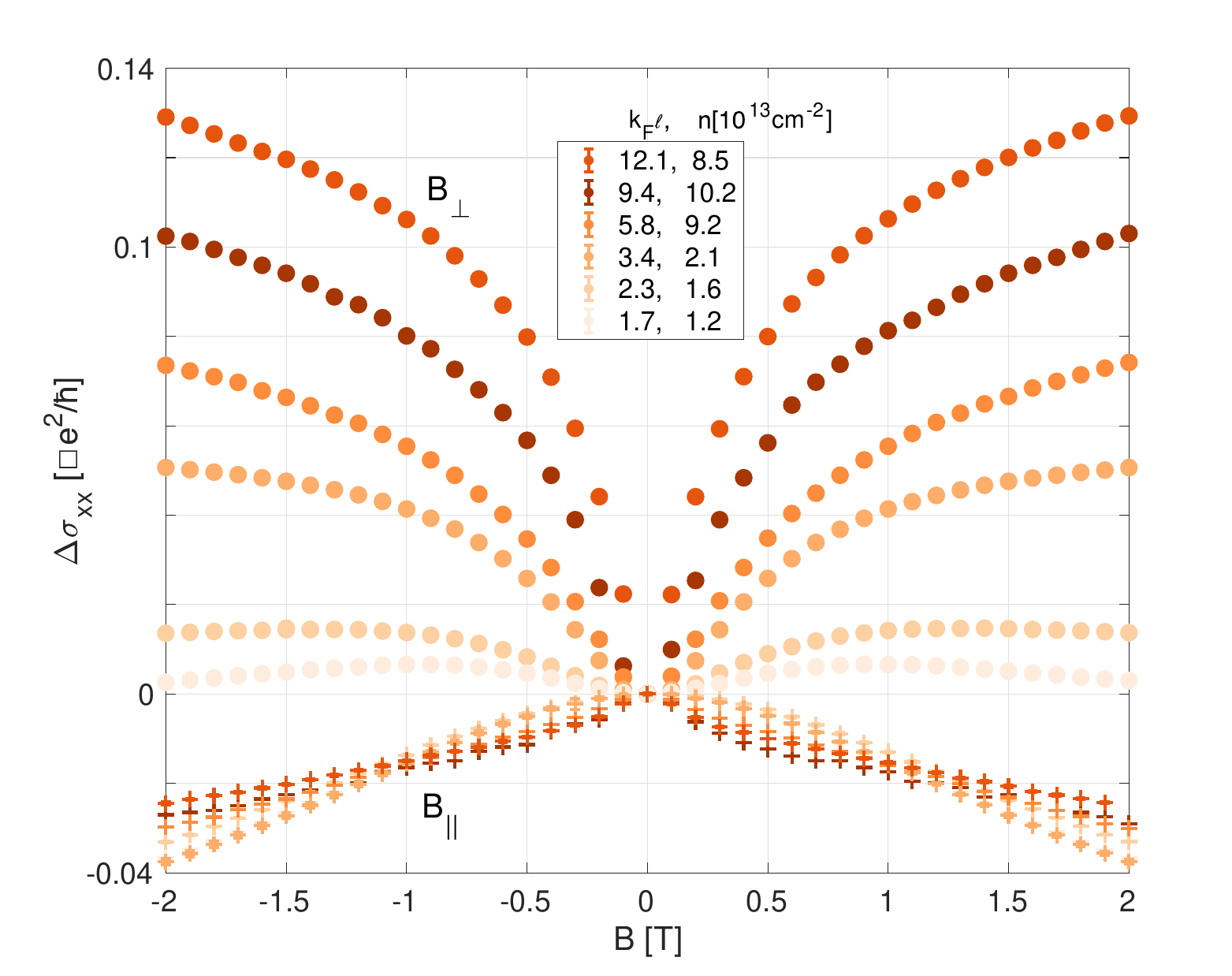}
  \caption{\label{fig_density_a}}  
  \end{subfigure}
\hfill   
  \begin{subfigure}{0.43\textwidth}  
  \includegraphics[width=\textwidth]{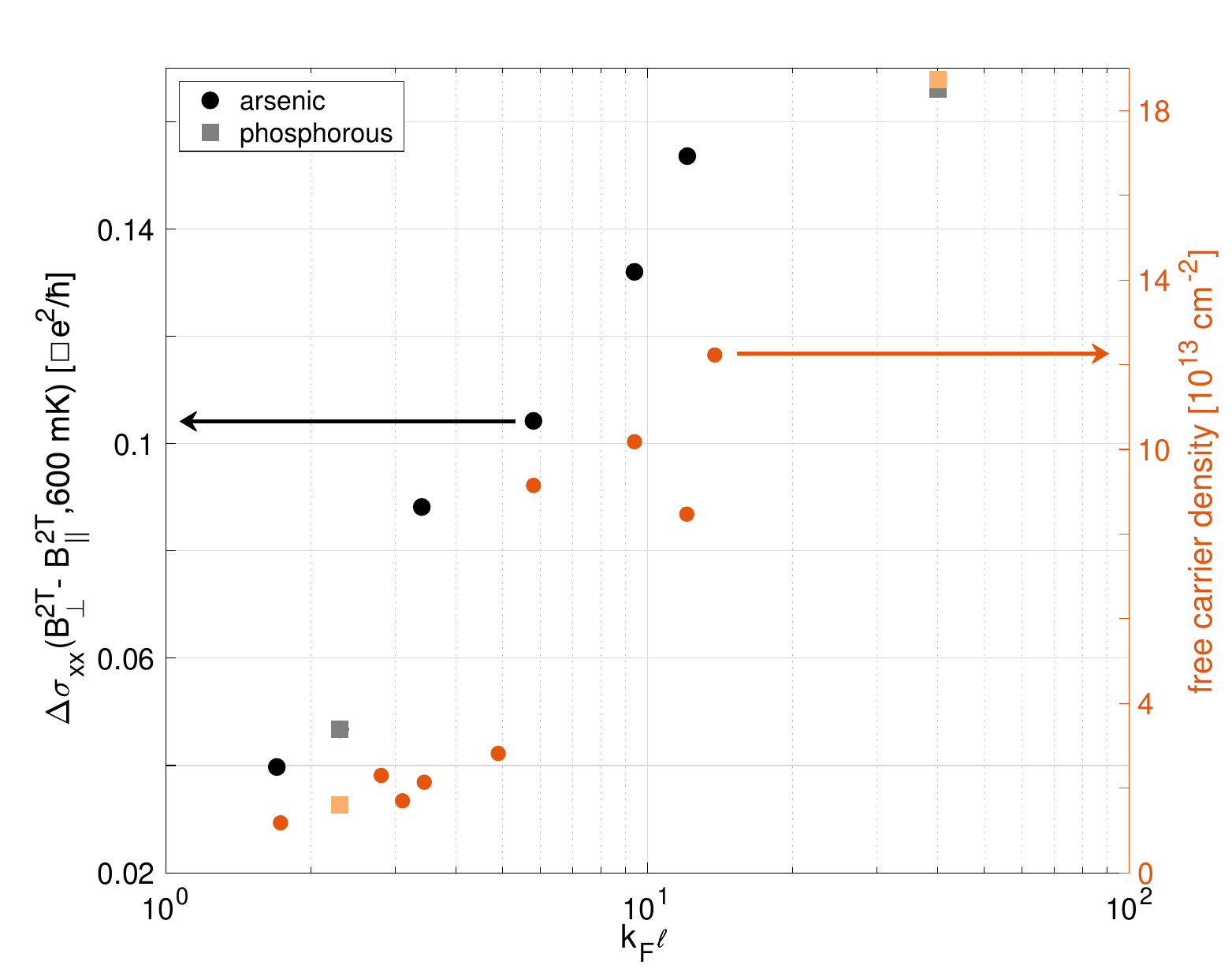}
  \caption{\label{MR_vs_kFL}}
  \end{subfigure}
     \caption{\textbf{Influence of dopant density on the magneto-conductance.} 
    (a) Magneto-conductance per square at 600~mK for samples with different free carrier densities $n_\text{2D}$. As $k_\text{F}\ell$  decreases with decreasing density the effect of the perpendicular magnetic field (full dots) becomes smaller. The conductivity for $B_\parallel$ is denoted by crosses.
	(b) 
    Magnitude of the weak-localization part of the magneto-conductance ($\Delta\sigma_{xx}(B_\perp)-\Delta\sigma_{xx}(B_\parallel)$ shown on the left $y$-axis) at 600~mK in a 2~T magnetic field as a function of $k_\text{F}\ell$. The right $y$-axis indicates the free carrier density. The out-of-plane magneto-conductance is seen to be dominated by weak localization corrections which depend on $k_\text{F}\ell$. 
      In (b) 
      the dots correspond to arsenic $\delta$-layers while the squares are from phosphorus $\delta$-layers, showing that the magneto-conductance does not depend on the dopant species.}     
    \label{fig_density}
\end{figure*}
\noindent\textbf{Electron-electron interactions in dilute $\delta$-layers}\\
Figure~\ref{fig_MR_a} shows in-plane and out-of-plane magneto-conductance {per square} obtained at 115~mK. At this temperature and sample density ($n = (1.18 \pm 0.01) \times 10^{13}$~cm$^{-2}$) the magneto-conductance is negative. This cannot be explained by weak localization, nor can the data be fit with weak anti-localization \cite{PhysRevB.53.3912}, nor with Minkov et al.'s model which includes Rashba and Dresselhaus spin-orbit interaction \cite{PhysRevB.70.155323, ILP_WAL, In_plane_InSb}, as could be expected for samples with large spin-orbit coupling. 
Instead, the magneto-conductance corrections are governed by two effects that add up in conduction as two channels in parallel: $\Delta\sigma_{xx}^{\rm tot} = \Delta\sigma_{xx}^{\rm WL} + \sigma_{xx}^{\rm Zeeman}$ \cite{PhysRevB.26.4009,PhysRevLett.47.1758}.
The first term $\Delta\sigma_{xx}^\text{WL}$ is the conventional positive weak localization as described in Eq.~(\ref{eq_HLN}) and Eq.~(\ref{eq_Sullivan}). The second term  $\Delta\sigma_{xx}^\text{Zeeman}$ has the opposite sign. It is the Zeeman effect in the Altshuler-Aronov (AA) \cite{AAcontribution} corrections to the conductivity as described in 1982 by Lee and Ramakrishnan \cite{PhysRevB.26.4009} and later elaborated on by Zala, Narozhny and Aleiner~\cite{PhysRevB.64.214204,zala2001interaction}:
\begin{equation}
\Delta\sigma(h) = -\frac{G_{0}}{2\pi}F\ln\left(\frac{h}{1.3}\right),\text{ for }h \gg 1.
\label{eq_Lee_Rama}
\end{equation}
Here $h = \frac{g\mu_\text{B}B}{k_\text{B}T}$ is the reduced magnetic field with  the $g$-factor $g=2$, $\mu_\text{B}$ the Bohr magneton and $k_\text{B}$ the Boltzmann constant, and $F$ is a dimensionless number (resulting from the angular average of the statically screened Coulomb scattering). The effect scales as $\ln(B/T)$, which becomes strong at low temperatures. 

The Zeeman effect results from the coupling of the magnetic field to the spin of the electrons, and thus (to the extent that spin orbit coupling does not interfere) is isotropic in the direction of the magnetic field, unlike weak localization that depends  mostly on the in-plane orbital motion and thus senses the perpenicular component of the magnetic field~\cite{2_to_3_crossover}. 
Thus, we can isolate the anisotropic WL contribution by taking the difference between the out-of-plane and in-plane magneto-conductance, as is done in Fig.~\ref{fig_MR_a} (black dots), and Fig.~\ref{WL_MR_vs_T}, where the full lines are fits with the weak-localization expression Eq.~(\ref{eq_HLN}) from which Eq.~(\ref{eq_Sullivan}) has been subtracted.
In Fig.~\ref{Greenford_vs_T} the orange points are the coherence lengths $\ell_\phi$ obtained from fitting the data at different temperatures, exhibiting an effective power law ($\ell_\phi\sim T^{-0.31}$) that reflects the decreasing rate of inelastic scattering \cite{PhysRevB.70.245423,2_to_3_crossover}. 
Note that when fitting these curves, only the coherence length $\ell_\phi$ and the thickness parameter $\gamma$ are free parameters. The mean free path $\ell$, the free carrier density and mobility are all obtained from Hall measurements at high temperatures (2~K), as shown in Fig.~\ref{fig_MR_a} in orange. In particular there is no room for a free pre-factor to Eq.~(\ref{eq_HLN}), even though it is often used with little or no justification for similar data \cite{PhysRevLett.112.236602, siP_sheet} (though for $\sigma_{xx} \lesssim10 G_0/\pi$ 
such a prefactor may effectively arise due to two-loop localization corrections \cite{Aleiner_1999, Minkov_alpha}). 
For thicker and denser samples, the out-of-plane magneto-conductance data could (erroneously) be fitted  to weak localization only, using a free pre-factor, because the AA correction is small. However, when subtracting out the AA term, no unjustified prefactor is needed.
From the fit parameters we can calculate the thickness of the dopant layer \cite{sullivan}, between 0.4 and 1.8~nm for the arsenic $\delta$-layers studied in this work, comparable to the silicon unit cell $\sim$0.5~nm. These are thinner than previously obtained layers in phosphorus doped Si \cite{doi:10.1063/1.4998712, PhysRevB.101.245419}, as is expected due to arsenic's slower diffusion \cite{Solmi_diffusivity} during annealing steps of sample fabrication, and corroborated by photoemission \cite{SiAs_ARPES} and X-ray reflectometry \cite{SiAs_XRR} experiments. It is interesting to note that in the thinnest layer only three Si(100) crystal planes are conductive, compared to thirteen in the thickest layer.

\begin{figure*}[ht]   
\centering
\begin{subfigure}{0.32\textwidth} 
    \includegraphics[width=\textwidth]{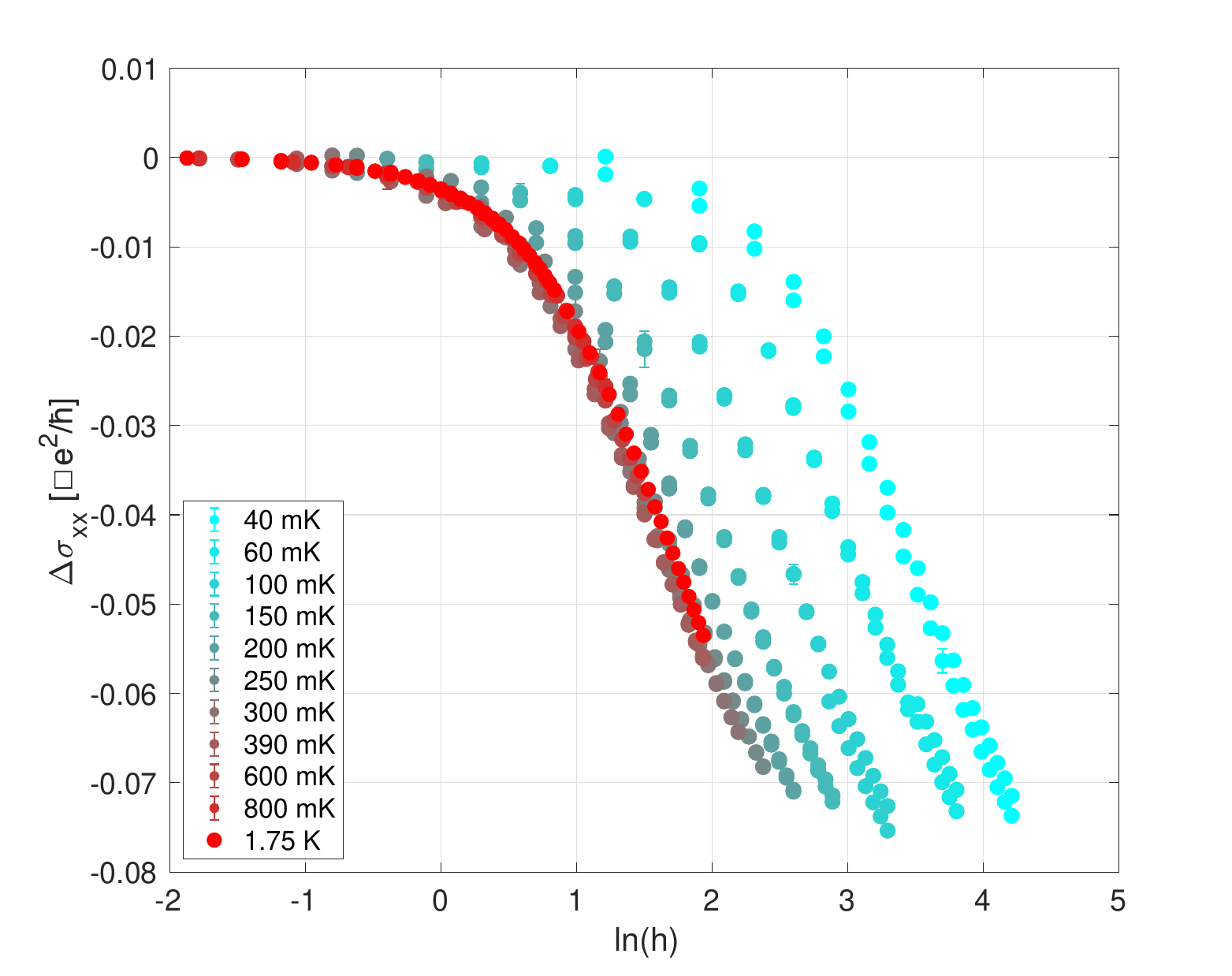}
    \caption{\label{MC_lnh}}
\end{subfigure}
\hfill 
\begin{subfigure}{0.32\textwidth} 
    \includegraphics[width=\textwidth]{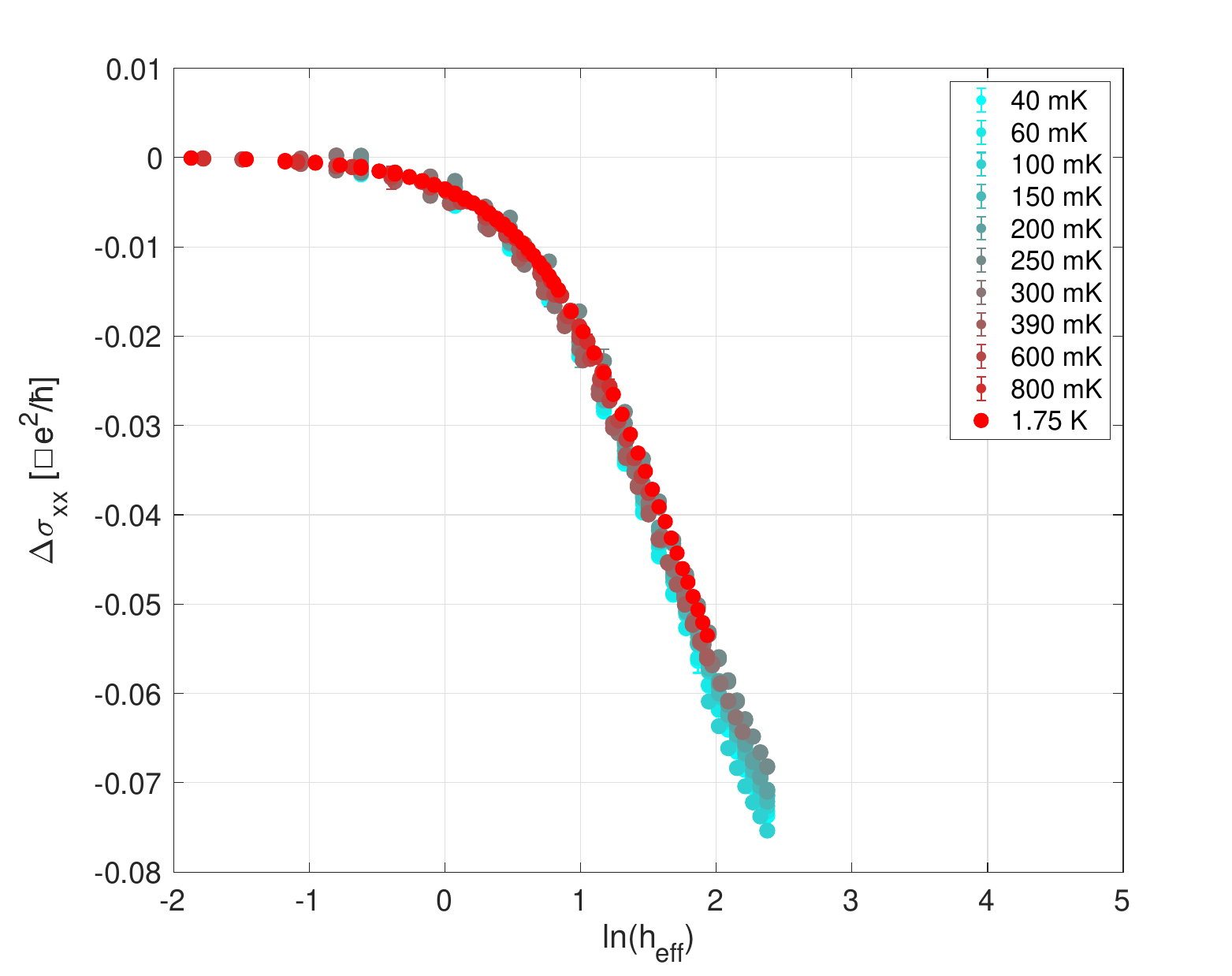}
    \caption{\label{MC_collapse}}
\end{subfigure}
\hfill 
\begin{subfigure}{0.32\textwidth} 
    \includegraphics[width=\textwidth]{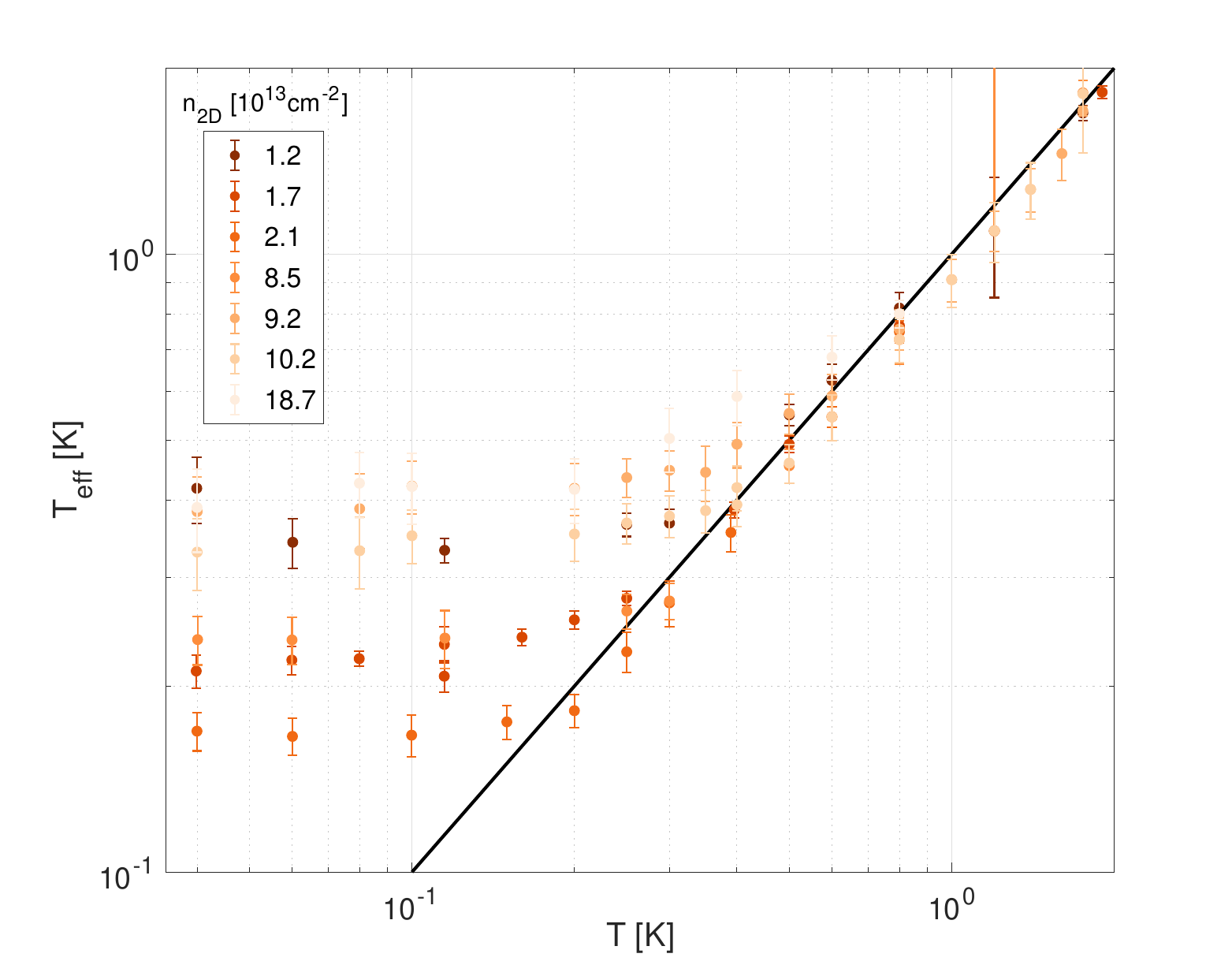}
    \caption{\label{T_collapse}}
\end{subfigure}
    \caption{\textbf{$B/T$ scaling of the AA correction.} (a) Change in conductivity due to the Zeeman splitting in units of the conductance quantum  $e^2/\hbar$, as a function of the logarithm of the reduced field $h=g\mu_\text{B} B/k_\text{B}T$. The data are shown for multiple temperatures and stem from a single sample with free carrier density $n = (2.14 \pm 0.02) \times 10^{13}$~cm$^{-2}$. The data collapse down to 250~mK.
      (b) Same as in (a), except that an effective temperature $T_{\rm eff}$ is used to obtain a full collapse of the data. $T_\text{eff}$ is plotted for all samples in (c).
      (c) Effective temperature $T_{\rm eff}$  determined for all samples so as to collapse the AA correction as in (a) and (b).}
\label{fig_zeeman}
\end{figure*} 

The weak localization effect calculated in Eq.~(\ref{eq_HLN}) stems from an expansion in $(k_\text{F}\ell)^{-1}$, where $k_\text{F}$ is the Fermi momentum of the electrons, which is proportional to the square root of the dopant density. For samples with different densities, the weak localization correction depends primarily on $k_\text{F}\ell$, which is consistent with the monotonic dependence of the perpendicular magneto-conductance as a function of $k_\text{F}\ell$, as shown in Fig.~\ref{fig_density_a} and Fig.~\ref{MR_vs_kFL} for $B= 2~T$ (see the supplemental section~\ref{sect_supplemental} for information on the variability of $\ell$ and $k_\text{F}$ across samples).
The electron-electron interaction effects depend on the ratio of Coulomb interaction and kinetic energy, which is proportional to the ionization energy divided by the hopping integral (the donors' wavefunction overlap) and thus is stronger for lower carrier densities. For in-plane magnetic fields, when the $\delta$-layer is thin and/or dilute enough, the electron-electron interaction contributions dominate over the weak localization term.  This is the case for our arsenic $\delta$-layers, as is visible in Fig.~\ref{fig_density_a}. This regime of thin and dilute half-filled layers has not been achieved in $\delta$-layers before.

To isolate the Zeeman terms in the Altshuler-Aronov corrections we remove the weak localization contribution to the magneto-conductance as determined by the fits shown in Fig.~\ref{fig_MR_a} and Fig.~\ref{WL_MR_vs_T}. 
The remaining electron-electron interaction contribution is given by Eq.~(\ref{eq_Lee_Rama}), which is a logarithmic function of the reduced field $h = \frac{g\mu_\text{B}B}{k_\text{B}T}$, and depends on $r_s$ through the average Coulomb scattering parameter $F$. Therefore, when the Zeeman contribution to the magneto-conductance is plotted against the reduced field $h$ at different temperatures, the data collapses to a single curve. This is shown in Fig.~\ref{MC_lnh} for a sample with density $n = (2.14 \pm 0.02) \times 10^{13}$~cm$^{-2}$, where a clear collapse of the data is seen down to 250~mK. This curve scales as $\ln(h)$ at large fields, as expected from Eq.~(\ref{eq_Lee_Rama}). In Fig.~\ref{MC_collapse} the same data are shown, but instead of the phonon temperature an effective electron temperature was used to obtain a collapse down to the cryostat base temperature of 40~mK. The effective temperatures used for the various samples to obtain the collapse are shown in Fig.~\ref{T_collapse} for all samples. The saturation of the effective electronic  temperatures at low bath temperature parallels the 
saturation of conductivity seen in the temperature dependence of the conductivity in Fig.~\ref{Greenford_vs_T}.

Devices in this study are mostly arsenic $\delta$-layers, except for one low density and one high density phosphorous $\delta$-layer (see Table.~\ref{table_samples}). The strength of the WL and AA corrections are summarized in Fig.~\ref{MR_vs_kFL} 
where results from the phosphorous $\delta$-layer are indicated as squares. While arsenic dopants are expected to have stronger spin-orbit coupling than phosphorous, due to their larger mass, we find that the WL and AA effects are independent of dopant species. This indicates that spin-orbit coupling has a negligible effect on magneto-conductance.
\\

\noindent\textbf{Conclusion}\\

In conclusion, magneto-transport measurements on $\delta$-layers of arsenic and phosphorous dopants in silicon show that the usual description of disordered metal conductivity by weak localization effects alone is insufficient in dilute and thin layers, due to significant electron-electron interaction effects. We find that when the conduction layers are sufficiently thin and dilute it becomes necessary to include the Zeeman splitting which diminishes the triplet enhancement of the conductivity~\cite{PhysRevB.26.4009,zala2001interaction}. We confirm that the effect of Zeeman splitting on the Altshuler-Aronov corrections is isotropic and combines additively with the weak-localization effect. 


\medskip
\textbf{Acknowledgments} \par 
 This project received funding from the European Research Council under the European Union’s Horizon 2020 research and innovation program, within the Hidden, Entangled and Resonating Order (HERO) project with Grant Agreement 810451. 
We acknowledge financial support of the Engineering and Physical Sciences Research Council (EPSRC) [grant numbers EP/R034540/1, EP/W000520/1]; and Innovate UK [75574]. J.B and P.C.C. were supported by the EPSRC Centre for Doctoral Training in Advanced Characterization of Materials (grant number EP/L015277/1), and by the Paul Scherrer Institut.
P.C.C. was partially supported by Microsoft.
M.M. was supported by SNSF Grant 200558.

\bibliographystyle{ieeetr}
\bibliography{biblio}	

\section{Supplemental information} \label{sect_supplemental}
\noindent\textbf{Experimental values for all measured samples}\\
Table~\ref{table_samples} contains the physical parameters of every sample, as obtained during fabrication and by fitting the magneto-conductance data.\\
\begin{table*}[hb!]
\centering
\begin{tabular}{cccc|cccccccc}
\# & Dopant & $d$ & RTA & $n_\text{2D}$ & $\mu$ & $\sigma_{xx}$ & $n_\text{2D}^{-1/2}$ & $\ell$ & $\ell_{\phi}$ & $\delta$ & $k_\text{F}\ell$\\
& species & (nm) & (°C) & (10$^{13}$ cm$^{-2}$) & (cm$^{2}$V$^{-1}$s$^{-1}$) & (10$^{-4}\square/\Omega$) & (nm) & (nm) & (nm) & (nm) & ( )\\ 
\hline
1 & P & 15 & 500 & 18.73$\pm$0.06 & 51.8$\pm$0.2 &  15.6$\pm$0.5 & 0.7 & 11.7$\pm$0.04 & 86$\pm$6 & 10$\pm$1 & 40.1$\pm$0.2\\ 
\hline
2 & As & 30 & 500 & 12.23$\pm$0.08 & 27.3$\pm$0.2 &  5.4$\pm$0.2 & 0.9 &  4.98$\pm$0.04& 75$\pm$1 & 1.46$\pm$0.08 & 13.8$\pm$0.1\\
\hline 
3 & As & 21 & 380 & 10.18$\pm$0.09 & 22.4$\pm$0.2 & 3.6$\pm$0.1 & 0.99 & 3.72$\pm$0.04 & 57.6$\pm$0.3 & 1.26$\pm$0.03 & 9.4$\pm$0.1\\
\hline 
4 & As & 21 & 250 & 9.15$\pm$0.09 & 15.4$\pm$0.2 &  2.25$\pm$0.07 & 1.05 & 2.42$\pm$0.03 & 41.9$\pm$0.2 & 0.41$\pm$0.04 & 5.80$\pm$0.09\\
\hline
5 & As & 30 & 500 & 8.47$\pm$0.03 & 34.4$\pm$0.1 & 4.7$\pm$0.1 & 1.1 & 5.23$\pm$0.02 & 77.4$\pm$0.5 & 1.04$\pm$0.04 & 12.07$\pm$0.06\\ 
\hline 
6 & As & 20 & 500 & 2.82$\pm$0.05 & 42.4$\pm$0.9 &  1.92$\pm$0.06 & 1.9 & 3.72$\pm$0.09 & 27.5$\pm$0.3 & 1.82$\pm$0.07 & 4.9$\pm$0.1\\ 
 \hline
7 & As & 30 & 500 & 2.14$\pm$0.02 & 38.9$\pm$0.3 &  1.33$\pm$0.04  & 2.2 &  2.97$\pm$0.02 & 35.5$\pm$0.2 & 0.42$\pm$0.06 & 3.44$\pm$0.04\\ 
 \hline
8 & As & 30 & 500 & 1.70$\pm$0.04 & 44.5$\pm$0.1 &  1.21$\pm$0.04 & 2.4 & 3.03$\pm$0.07 & 23.2$\pm$0.2 & 0.88$\pm$0.06 & 3.1$\pm$0.1\\ 
\hline
9 & P & 15 & 500 & 1.61$\pm$0.02 & 34.5$\pm$0.5 &  0.89$\pm$0.03  &  2.5 & 2.29$\pm$0.04 & 20.5$\pm$0.5 & 1.3$\pm$0.1 & 2.30$\pm$0.05\\ 
 \hline
10 & As & 30 & 500 & 1.18$\pm$0.01 & 35.6$\pm$0.3 &  0.67$\pm$0.02 & 2.9 & 2.01$\pm$0.02 & 22.4$\pm$0.2 & 0.65$\pm$0.05 & 1.73$\pm$0.02\\ 
  \hline   \end{tabular}
\caption{\textbf{Characteristics of the electron layers in this work.} The depths $d$ of the layers are deduced from secondary ion mass spectrometry. \textquote{RTA} refers to the final temperature of the rapid thermal annealing  during fabrication. All other values are extracted from magneto-conductance measurements taken at $T>2$~K. $n_\text{2D}$ is the 2D electron density, $\mu$ the electron mobility, $\sigma_{xx}$ the conductivity at 2~K, $n_\text{2D}^{-1/2}$ the average distance between donors in 2D, $\ell$ the electron mean free path, $\ell_\phi$ the electron coherence length, $\delta$ the thickness of the $\delta$-layer , and $k_\text{F}$ the Fermi wave-vector.}
\label{table_samples}
\end{table*}

\noindent\textbf{Sample fabrication}\\ 
Ten $2\times9$~mm Si(001) samples were diced from Czochralski-grown wafers (bulk doping~$<5\times10^{14}$~cm$^{-3}$), and cleaned ultrasonically in acetone, followed by isopropyl alcohol. Each sample was thermally outgassed in vacuum (base pressure~\mbox{$<5\times10^{-10}$~mbar}) for~\textgreater~8~h at 600~°C, then flash annealed multiple times at 1200~°C using direct current resistive sample heating. This temperature was monitored using an infrared pyrometer (\mbox{IMPAC IGA50-LO plus}) with an uncertainty of $\pm30~^{\circ}$C.
Each sample was dosed with AsH$_3$ or PH$_3$ for a different duration to vary the dopant density. 
The sample was then heated to 350~°C for $2-5$~min 
to incorporate the arsenic into the silicon lattice \cite{Goh2005}. Afterwards, a $1-2$~nm silicon \textquote{locking layer} was deposited, without resistive sample heating, to confine the dopants \cite{Keizer2015, Stock_As_count}. A 15~s \textquote{rapid thermal anneal} at 500~°C was performed for eight of the samples, to improve electrical activation. Sample 4 was not heated, and Sample 3 was heated to~380~°C. 
More silicon ($14-28$~nm depending on sample) was deposited, with the sample at 250~°C. Silicon was deposited at a rate of $0.1-0.2$~nm/min using a silicon solid sublimation source (\mbox{SUSI-40}, MBE Komponenten GmbH). During deposition, the sample temperature was indirectly monitored by the sample resistance, while heating using a direct current resistive sample heater. Deposition rates and depths are estimated from secondary ion mass spectrometry measurements.

The δ-layers were then processed into $20\times200$~μm Hall bars using optical lithography (Quintel Q4000-6 Mask Aligner) and reactive ion etching (Oxford Instruments Plasma Pro NGP80 RIE) to remove the δ-layer outside the bar and to etch holes for aluminium contacts \cite{Fuhrer2009}.
These contacts were deposited using optical lithography followed by HF etching to remove the surface oxide, immediately before Al sputter deposition (Kurt Lesker PVD75).\\

\noindent\textbf{$B/T$ scaling of the AA correction.}\\
In Fig.~\ref{MR_vs_kFL} on the right y-axis, $k_\text{F}\ell$ is shown as a function of sample density. Note that if the roughness and disorder of the dopant layers do not vary much from sample to sample, then $k_\text{F}\ell$ increases monotonically with the density and it looks like the weak-localization effect is governed by density rather than $k_\text{F}\ell$, as was also observed in the case of the experiment on Si:P  \cite{siP_sheet}. In our samples, however, $k_\text{F}\ell$ is not a monotonic function of the density, likely because of variable disorder and layer thickness, allowing us to distinguish the effects that scale with the density (the AA correction) from the effects that scale with $k_\text{F}\ell$ (the weak-localization effect). 

The Fig.~\ref{MR_vs_kFL} 
 contains data from phosphorus $\delta$-layers (squares), showing that the effects depend predominantly on the dopant density or $k_\text{F}\ell$, but not on the dopant species.

\end{document}